# Prediction of the Optimal Threshold Value in DF Relay Selection Schemes Based on Artificial Neural Networks

[1]Ferdi KARA, [1]Hakan KAYA, [2]Okan ERKAYMAZ, [1]Ertan ÖZTÜRK
[1]Department of Electrical Electronics Engineering, [2]Department of Computer Engineering
Bulent Ecevit University,
Zonguldak, Turkey
{f.kara@beun.edu.tr, hakan.kaya@beun.edu.tr, okan.erkaymaz@beun.edu.tr, eozturk@beun.edu.tr}

*Abstract*—In wireless communications, the cooperative communication (CC) technology promises performance gains compared to traditional Single-Input Single Output (SISO) techniques. Therefore, the CC technique is one of the nominees for 5G networks. In the Decode-and-Forward (DF) relaying scheme which is one of the CC techniques, determination of the threshold value at the relay has a key role for the system performance and power usage. In this paper, we propose prediction of the optimal threshold values for the best relay selection scheme in cooperative communications, based on Artificial Neural Networks (ANNs) for the first time in literature. The average link qualities and number of relays have been used as inputs in the prediction of optimal threshold values using Artificial Neural Networks (ANNs): Multi-Layer Perceptron (MLP) and Radial Basis Function (RBF) networks. The MLP network has better performance from the RBF network on the prediction of optimal threshold value when the same number of neurons is used at the hidden layer for both networks. Besides, the optimal threshold values obtained using ANNs are verified by the optimal threshold values obtained numerically using the closed form expression derived for the system. The results show that the optimal threshold values obtained by ANNs on the best relay selection scheme provide a minimum Bit-Error-Rate (BER) because of the reduction of the probability that error propagation may occur. Also, for the same BER performance goal, prediction of optimal threshold values provides 2dB less power usage, which is great gain in terms of green communication.

*Keywords—relay selection, cooperative, ANNs, MLP, RBF, optimal threshold*

## I. INTRODUCTION

The performances of wireless communication techniques are mostly evaluated according to bit-error rate (BER) and outage probability. These two criteria's depend on the wireless channel properties such as interference, fading, shadowing and path-loss etc[1]. Researchers have been studying on several techniques in order to increase the performance. One of these techniques for improving the performance is diversity in which the copies of data are transmitted over independent dimensions such as time, frequency and space (antennas). In spatial diversity, the copies of data are transmitted/received by multiple antennas. Therefore, the spatial diversity is mostly called as Multiple Input Multiple Output (MIMO) system [2], which is key technology in 4G standards i,e LTE and LTE-Advanced [3]. However, in mobile communications multiple antennas cannot be easily adapted due to physical limitations. To overcome these limitations the relaying systems are proposed in literature. In the relaying systems, the other users -called relays- in the environment forward a processed version of the received data from source to the destination. The relaying systems are seen the nominees for the 5G and beyond technologies. This relaying system is called cooperative communication (CC) and the obtain diversity is called cooperative diversity or virtual spatial diversity [4]. In literature, different relaying protocols such as amplify-and-forward (AF) and decode-and-forward (DF) are used according to the employed processing at the relay. In AF protocol, the received data is amplified by a relay (to equalize the effect of the channel fade between the source and the relay) then the relay retransmits the amplified version of data to a destination [5]. On the other hand, in DF protocol, a relay first decodes the received data, then re-encodes and forwards to a destination. In the DF protocol, the relay decides whether it transmits or not according to the source-relay link quality. If the source-relay link quality is less than a threshold value on the relay, relay remains silent. Otherwise the relay transmits [6]. However, it is still possible that the relay may decode incorrectly even if a threshold value is greater than the source-relay link quality and decode the data from the source and then forward this incorrect data to the destination erroneously. This phenomenon is called error propagation problem. The probability of this incorrect detection at the relay depends on the threshold value [7].

In a CC system, the source could transmit the signal to the destination not only through the one relay but also through multi relays (all available relays) to increase diversity order gain .Instead of using all available relays, only one relay, which is the best one among the multi relays to obtain best performance, can be selected. The selected relay has the highest link qualities in terms of the transmission path and called best relay. Hence this scheme is called relay selection [8].The use of only the best relay also provides the full diversity order with an efficient use of the bandwidth.

In the DF protocol and relay selection scheme, the threshold value has great effect on the performance of cooperative communication. To minimize BER, an optimal threshold values has to be obtained. This optimal threshold value depends on the number of relays and the link qualities

This work has been supported by The Scientific and Technological Research Council of Turkey (TUBITAK) with the name 2211A PhD Scholarship Program.

between source-relay, relay-destination and source-destination [7]. The computation of optimal threshold value changing relay numbers and link qualities by analytically is hardly possible.

In this paper, for the prediction of optimal threshold values by using MLP and RBF networks, the number of relays and the average link qualities of source-relay-destination paths are used as inputs after normalization.

The optimal threshold values on the best relay selection scheme are determined for different scenarios by numerically minimizing the closed form BER equation similar to given in [6]. Different types of scenarios having different relay numbers and different link qualities have been used to test the proposed techniques. The outputs of ANNs are perfect match with the numerical results.

This paper organized into six sections. In section II, we have briefly explained the best relay selection in DF network models. Also the numerical results for optimal threshold values are given in this section. Section 3, has an overview of literature in determination of threshold values on DF networks. Section 4 deals with the description of ANNs as proposed techniques. Application of ANNs for the prediction of optimal threshold values in DF networks and results are given in Section 5. In section 6, some results are discussed, and the paper is concluded.

## II. SYSTEM MODEL

In the relay selection scheme, one relay is selected among multi-relays in order to utilize full cooperative diversity [8]. The relay selection is performed by considering source-relay-destination cascaded link qualities of all available relays. Relay selection is done in two steps: First, the decoding set (C) of relays which are called reliable relays, is determined by considering source-relay link qualities of all available relays. If the received SNR from source at the relay is less than a threshold value, the relay cannot decode the source message correctly and it does not transmit. If the received SNR at the relay is greater than the threshold, this relay is added into the decoding set. It is also assumed that the relays belong to the decoding set may detect the signals from source erroneously and forward incorrect data to the destination, causing error propagation that can reduce the performance of the system [9]. Second, the relay selection is completed at the destination by taking the relay having the highest relay-destination link quality from the decoding set. After determining the best relay, the destination informs all relays about which relay is selected to transmit as the best relay through a reverse broadcast channel then other unselected relays turn to be idle [10]. The destination combines the data received from source and best relay by using Maximum Ratio Combining (MRC). The best relay selection scheme is given Fig 1.

For the best relay selection scheme, the BER of system, which is given in the (1), is obtained by using the equations given in [5] and [11] under the condition that the destination uses MRC and the modulation is Binary Phase Shift Keying (BPSK). The BER of systems depends on the threshold value ($\gamma_{th}$), the average link SNRs ($\overline{\gamma_{SR}}, \overline{\gamma_{RD}}, \overline{\gamma_{RD}}$) and the number of relays (M).

The definition of erfc(.) function used in (1) is $erfc(x) = \frac{2}{\pi}\int_0^{\pi/2} exp(-\frac{x^2}{sin^2(\theta)})d_\theta$.

The optimal threshold ($\gamma_{th}^{opt}$) is the value which minimizing the BER of the system.

$$\gamma_{th}^{opt} = arg, min\{\gamma_{th}, BER\} \quad (2)$$

In our scenarios, it is assumed that the average source-relay, relay-destination and source-destination link SNRs are $\bar{\gamma}_{SR} = E(h_{SR}^2) \cdot \varepsilon_b/N_0$, $\bar{\gamma}_{RD} = E(h_{RD}^2) \cdot \varepsilon_b/N_0$, $\bar{\gamma}_{SD} = E(h_{SD}^2) \cdot \varepsilon_b/N_0$ respectively. E(.) is the statistical average (or expectation) operator. $h_{SD}$, $h_{SR}$ and $h_{RD}$ are the channel coefficients and they are modeled as Rayleigh flat fading channels with variance of $\sigma_{SD}^2$, $\sigma_{SR}^2$ and $\sigma_{RD}^2$, respectively. In Fig. 2, the optimal threshold value ($\gamma_{th}^{opt}$) numerically obtained minimizing BER as a function of $\varepsilon_b/N_0$ is showed for different number of relays (M) on the symmetric network which has $\sigma_{SD}^2 = \sigma_{SR}^2 = \sigma_{RD}^2 = 1$.

In Fig. 3, the optimal threshold value ($\gamma_{th}^{opt}$) is showed on another network in which $\sigma_{SR}^2 = \sigma_{RD}^2 = 10$ and $\sigma_{SD}^2 = 1$. It is mostly possible scenario if the relay stands between source and destination.

$$BER = \left(1 - \exp\left(-\gamma_{th}/\overline{\gamma_{SR}}\right)\right)^M \times 0.5\, erfc\left(\sqrt{\overline{\gamma_{SD}}}\right) + \sum_{i=1}^{M}\left[\exp\left(-\gamma_{th}/\overline{\gamma_{SR}}\right)^i \left(1 - \exp\left(-\gamma_{th}/\overline{\gamma_{SR}}\right)\right)^{M-i} \left[\left(\frac{\overline{\gamma_{RD}}}{\overline{\gamma_{RD}}+\overline{\gamma_{SD}}}\right) \times \right.\right.$$
$$\left\{0.5\left(erfc(\sqrt{\gamma_{th}}) - \exp\left(\gamma_{th}/\overline{\gamma_{SR}}\right)\sqrt{\frac{\overline{\gamma_{SR}}}{1+\overline{\gamma_{SR}}}}\, erfc\left(\sqrt{\gamma_{th}\left(1+\frac{1}{\overline{\gamma_{SR}}}\right)}\right)\right)\right\} + \left\{1 - 0.5\left(erfc(\sqrt{\gamma_{th}}) - \exp\left(\gamma_{th}/\overline{\gamma_{SR}}\right)\sqrt{\frac{\overline{\gamma_{SR}}}{1+\overline{\gamma_{SR}}}}\times\right.\right.$$
$$\left.\left.erfc\left(\sqrt{\gamma_{th}\left(1+\frac{1}{\overline{\gamma_{SR}}}\right)}\right)\right)\right\} \times 0.5\left\{M\sum_{k=0}^{M-1}\frac{(-1)^k}{k+1}\binom{M-1}{k}\times\left(1 - \frac{\overline{\gamma_{RD}}/k+1}{\overline{\gamma_{RD}}/k+1-\overline{\gamma_{SD}}}\sqrt{\frac{\overline{\gamma_{RD}}/k+1}{1+\overline{\gamma_{RD}}/k+1}} + \frac{\overline{\gamma_{SD}}}{\overline{\gamma_{RD}}/k+1-\overline{\gamma_{SD}}}\sqrt{\frac{\overline{\gamma_{SD}}}{1+\overline{\gamma_{SD}}}}\right)\right\}\right] \quad (1)$$

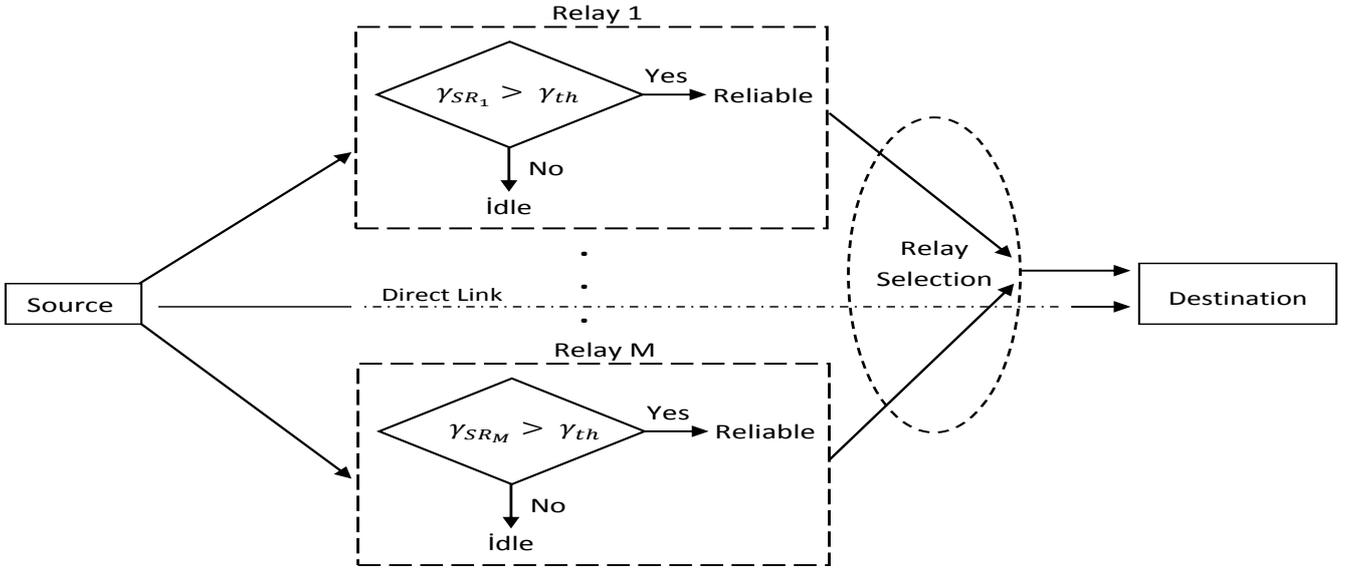

Figure 1. Best Relay Selection Scheme

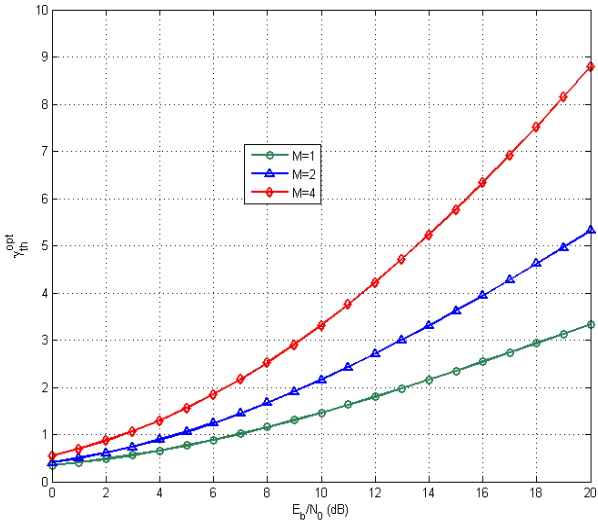

Figure 2. The optimal threshold values versus $\varepsilon_b/N_0$ on the symmetric network $\sigma_{SD}^2 = \sigma_{SR}^2 = \sigma_{RD}^2 = 1$ for M=4

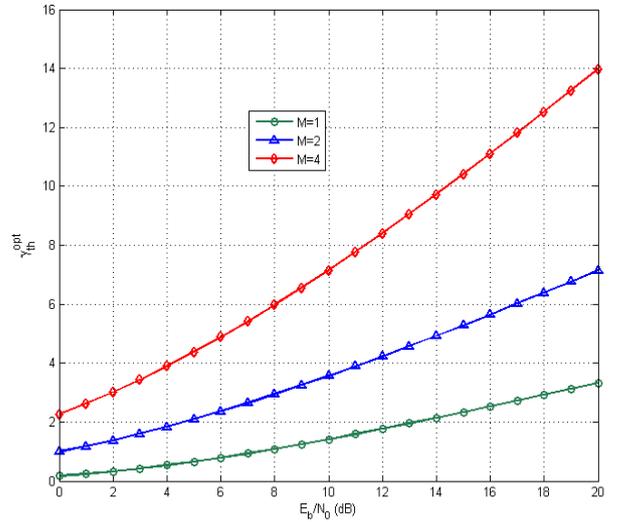

Figure 3. The optimal threshold values versus $\varepsilon_b/N_0$ on the network $\sigma_{SR}^2 = \sigma_{RD}^2 = 10$ and $\sigma_{SD}^2 = 1$ for M=4

## III. RELATED WORKS

The optimal threshold value in DF schemes could have been obtained by analytically just for a single relay i,e M=1 over Rayleigh fading channels [7] and over Nakagami-m fading channels [12] by using optimum decision rule similar to given (2). Similarly in [13], the authors proposed a method for the optimal threshold value and optimal power allocation when single relay is used. In [7], the authors have obtained the optimal values numerically by minimizing BER values as described Section II for the multiple relay scenarios. However, this method could not be used at the relay because of the computational burden. In [14], the authors used the MATLAB Optimization Toolbox for the determination of the threshold values.

## IV. OVERVIEW OF ARTIFICIAL NEURAL NETWORKS

The Artificial Neural Networks (ANNs) have been studied by researchers since 1970s. Artificial Neural Networks (ANNs) have become commonly used tool to generate proper outputs for given inputs in the absence of a formula between inputs and outputs. An ANN model is formed by input(s) with weight(s), activation function(s), biases an output(s) [15].The weights and biases are changed by the time an output is generated with a tolerated error for given inputs. Thanks to their computational speed, ability to handle complex functions and great efficiency even if full information is absent for the problem, ANNs have become one of the most preferred methods to solve non-linear engineering problems. ANNs are mainly used for classification, function approximation, clustering and regression [16]. ANNs have been also used in wireless communications for the purpose of channel estimations, channel equalizations etc [17] [18] [19]. In the last years, researchers have started to use the ANNs for the network solutions as relay selection methodology[3].

ANNs have different types which are called according to network connections and the activation functions used. Multi-Layer Perceptron (MLP) and Radial Basis Function (RBF) networks are two of the most popular ANNs.

### A. Multi-Layer Perceptron (MLP) Network

Multi-Layer Perceptron (MLP) network which is proper for the linear and non-linear applications is the most used ANN network. MLP network consists of an input layer, one or more hidden layers and an output layer. A MLP network with one hidden layer is shown Fig. 4.

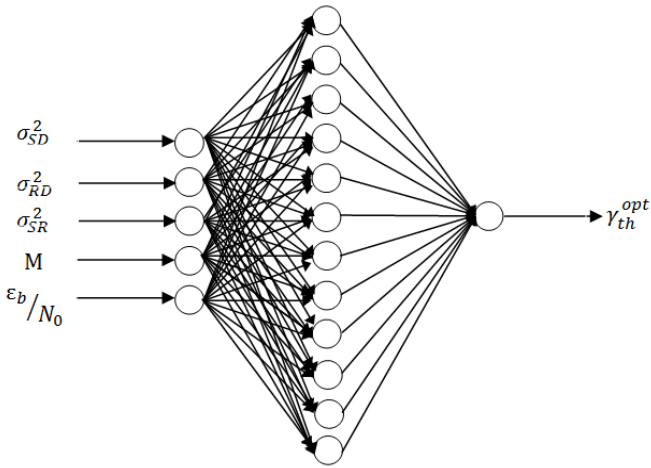

Figure 4. Multi-Layer Perceptron Network with one hidden layer

The computations are located on the neurons at the each layer and the information is transferred forward from node to node via weighted connections. The weights on the connections and the biases values for each neuron are adjusted according to desired outputs via backpropagation algorithms [20]. There are several learning algorithms in the literature. In this work, one of the mostly used algorithm Levenberg-Marquardt back propagation is preferred. MLP is chosen in this work due to its simple structure for prediction problems and its efficiency in learning large data sets. The output of MLP with one hidden layer is:

$$y = f_o\{\sum_{k=1}^{m} f_h(\sum_{j=1}^{n} w_{hj} x_j + b_{hj}) w_{ok} + b_{ok}\}. \quad (3)$$

In (3) $x_j$ is input(s) and $y$ is the output. $f_o$, $w_{hj}$, $b_{hj}$ are hidden layer activation function, input-hidden layer weights and hidden layer biases, respectively. Likewise $f_h$, $w_{ok}$, $b_{ok}$ are output layer activation function, hidden-output layer weights and output layer biases, respectively. In this work "tangent sigmoid" and "pure linear" functions are used as a hidden layer activation and output layer activation function.

### B. Radial Basis Function (RBF) Network

Radial Basis Function (RBF) network also consists of three layers: input layer, hidden layer and output layer like MLPs. However, RBF networks are trained faster than MLPs. The RBF networks could be defined as an application of neural networks for function approximation on the multi dimension space. RBF networks use Gauss function as hidden layer activation function. The output of a RBF network is:

$$y = \sum_{k=1}^{n} \left( e^{-\left\|\frac{x_j - \mu_j}{2\sigma^2}\right\|} \right) w_{ok} + b_{ok}. \quad (4)$$

In (4) $x_j$ is the input and $y$ is output of network. $\mu_j$ $\sigma^2$ are neurons center points and spread, respectively. The output of the hidden layer is transferred to output after multiplied by hidden-output layer weights ($w_{ok}$) and summed with output biases ($b_{ok}$). In the RBF network the input-hidden layer weights are all 1 [20].

## V. PREDICTION OF OPTIMAL THRESHOLD VALUES

The BER values for the selection relaying scheme given in (1) are calculated for various threshold values on the different scenarios having changing number of relays and link qualities as described Part II. Then, the threshold value which minimizes the BER is assumed the optimal. The data set explained above how to obtain is divided into two parts for training and testing. The training data has 12500 samples and the testing data has 3125 samples. After normalized, the training data is applied to ANNs: MLP and RBF networks, in the same way.

During the training of MLP, the network was trained many times for different numbers of hidden layer neurons, and the performance criteria –Mean Squared Error (MSE) - was noted. The more neurons in the hidden layer, the less MSE is obtained on training. However, the changing of MSE after 12 neurons is very slow. In addition to that, considering the complexity to implement of MLP, the number of hidden layer neurons is chosen 12. The training MSEs of MLP according to different number of hidden neurons are given Table 1.

TABLE I. THE TRAIN MSEs OF MLP ACCORDING TO NUMBER OF HIDDEN LAYER NEURONS

| Number of Hidden Layer Neurons | MSEs |
|---|---|
| 4 | 2.62 E-04 |
| 6 | 8.49 E-05 |
| 8 | 1.45 E-05 |
| 10 | 1.16 E-05 |
| 12 | 8.59 E-06 |
| 14 | 2.53 E-06 |
| 16 | 1.58 E-06 |
| 18 | 1.11 E-06 |

The RBF network was trained with the same number of neurons to compare with MLP network properly. The spread of the network on RBF has effect on training MSE. However, there is not a way to calculate the spread coefficient. Therefore, the RBF network was retrained for different spread coefficients until the minimum MSE is obtained. The spread coefficient is chosen 0.8 for RBF network.

After training of two networks, they were tested with dataset not used for training. For the 10 different samples randomly selected from the test data, the outputs of ANNs are obtained. Three statistical performance criteria's –Mean Squared Error (MSE), Mean Absolute Error (MAE) and coefficient of determination ($R^2$)- are calculated to compare the ANNs outputs with the numerical optimal threshold values. The results are given Table 2.

TABLE II. THE COMPARISON OF NUMERICAL RESULTS WITH ANNs OUTPUTS AND THE STATISTICAL PERFORMANCE CRITERIA'S OF ANNs FOR 10 SAMPLES RANDOMLY SELECTED

| S. No | Inputs | | | | | Num. Values | ANN Outputs | |
|---|---|---|---|---|---|---|---|---|
| | $\sigma_{SR}^2$ | $\sigma_{RD}^2$ | $\sigma_{SD}^2$ | $M$ | $\varepsilon_b/N_0$ | $\gamma_{th}^{opt}$ | MLP | RBF |
| 1 | 1 | 5.5 | 10 | 2 | 9 | 0.1100 | 0.1094 | 0.989 |
| 2 | 10 | 3.25 | 10 | 8 | 8 | 0.3432 | 0.3371 | 0.3377 |
| 3 | 1 | 3.25 | 7.75 | 4 | 0 | 0.3171 | 0.3222 | 0.3056 |
| 4 | 3.25 | 5.5 | 7.75 | 6 | 0 | 0.0753 | 0.0786 | 0.0791 |
| 5 | 3.25 | 1 | 10 | 2 | 3 | 0.0750 | 0.0739 | 0.0652 |
| 6 | 10 | 10 | 10 | 6 | 13 | 0.4438 | 0.4446 | 0.4387 |
| 7 | 7.75 | 3.25 | 3.25 | 4 | 20 | 0.4291 | 0.4303 | 0.4190 |
| 8 | 3.25 | 5.5 | 3.25 | 8 | 10 | 0.2839 | 0.2805 | 0.2861 |
| 9 | 1 | 7.75 | 5.5 | 8 | 18 | 0.4107 | 0.4090 | 0.4540 |
| 10 | 7.75 | 3.25 | 7.75 | 8 | 16 | 0.6127 | 0.6125 | 0.6196 |
| | | | | | | MSE | 9.225 E-06 | 2.4516 E-04 |
| | | | | | | MAE | 0.0024 | 0.0109 |
| | | | | | | $R^2$ | 0.9997 | 0.9914 |

On the two network scenarios described in Part II, the ANNs outputs are calculated like given in (3) and (4). In Fig. 5, on the symmetric network -$\sigma_{SD}^2 = \sigma_{SR}^2 = \sigma_{RD}^2 = 1$- the optimal values numerically obtained and the ANNs outputs are given according to $\varepsilon_b/N_0$ values for M=4. In Fig. 6, the same computations are given for M=6 on the other network - $\sigma_{SR}^2 = \sigma_{RD}^2 = 10$ and $\sigma_{SD}^2 = 1$-.

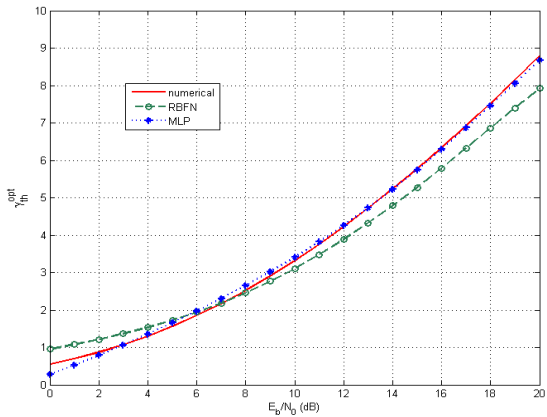

Figure 5. The numerical optimal thresholds and ANNs' outputs versus $\varepsilon_b/N_0$ for M=4 on the symmetric network in which $\sigma_{SD}^2 = \sigma_{SR}^2 = \sigma_{RD}^2 = 1$

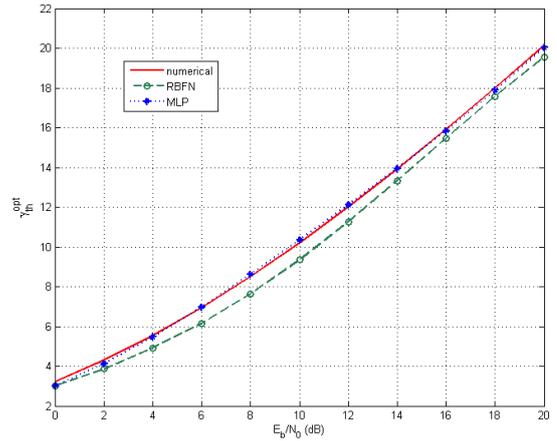

Figure 6. The numerical optimal thresholds and ANNs' outputs versus $\varepsilon_b/N_0$ for M=6 on the network in which $\sigma_{SR}^2 = \sigma_{RD}^2 = 10$ and $\sigma_{SD}^2 = 1$.

## VI. CONCLUSION

We have proposed the use of two different ANN models (MLP and RBF) for the first time to obtain the optimal threshold values for the best relay selection scheme in cooperative communications systems. Different scenarios having different link qualities and different number of relays were used to test the validity of the proposed models. Numerical results have shown that the optimal threshold value increases as a function of relays number and of SNR which represents all link qualities. The results have shown that two ANN networks can predict the optimal values. However, the MLP network outperforms the RBF network when using the same number of hidden layer neurons.

By the prediction of the optimal threshold value at the relay adaptively, BER value of the DF scheme remain minimum all SNR values. Otherwise, small threshold values would have low BER performance at high $\varepsilon_b/N_0$ values because of the propagation error; high threshold values would have low BER performance at the low $\varepsilon_b/N_0$ value. On the symmetric network -$\sigma_{SD}^2 = \sigma_{SR}^2 = \sigma_{RD}^2 = 1$- which has M=4 relays, the BER performances of the system have obtained for different constant threshold values and for the optimal threshold values according to changing $\varepsilon_b/N_0$ values.

TABLE III. THE COMPARISON OF THE BER VALUES FOR ACHIVED BY OPTIMAL THRESHOLD VALUES WITH THE ACHIVED BY CONSTANT THRESHOLD VALUES ON TH SYMETRIC NEWROKS WITH 4 RELAYS

| Threshold Values | $\varepsilon_b/N_0$ (dB) | | | | | | |
|---|---|---|---|---|---|---|---|
| | 0 | 4 | 8 | 10 | 12 | 16 | 20 |
| 1 | 7,32 E-02 | 1,54 E-02 | 4,74 E-03 | 3 E-03 | 1,93 E-03 | 7,92 E-04 | 3,19 E-04 |
| 3 | 0,130 | 2,8 E-02 | 2 E-03 | 5,55 E-04 | 2,29 E-04 | 7,96 E-05 | 3,19 E-05 |
| 5 | 0,144 | 4,96 E-02 | 4,67 E-03 | 9,26 E-04 | 1,61 E-04 | 1,12 E-05 | 3,63 E-06 |
| 10 | 0,147 | 7,31 E-02 | 1,58 E-02 | 4,41 E-03 | 9 E-04 | 1,99 E-05 | 2,96 E-07 |
| $\gamma_{th}^{opt}$ (MLP) | 6,98 E-02 | 1,48 E-02 | 1,86 E-03 | 5,39 E-04 | 1,38 E-04 | 6,83 E-06 | 2,44 E-07 |
| $\gamma_{th}^{opt}$ (RBF) | 7,2 E-02 | 1,51 E-03 | 1,85 E-03 | 5,45 E-04 | 1,43 E-04 | 7,38 E-06 | 2,87 E-07 |

In summary, obtaining the threshold value adaptively with the ANNs for average $\varepsilon_b/N_0$ values and number of relays provides minimum BER values for all $\varepsilon_b/N_0$ values. The results also shown that, compared to constant threshold values using optimal threshold values at the relay gives us 2 dB less power opportunity for the same BER performances. This is very promising result in terms of green communication systems.

As a conclusion, this paper shows that by the time the threshold value is obtained properly, the DF cooperation communication systems are very promising for 5G standards instead of the traditional MIMO techniques already exist in 4G standards